# A deterministic model for forecasting long-term solar activity

Eleni Petrakou

*Athens, Greece*
*eleni@petrakou.net*

**Abstract**

A phenomenological model is presented for the quantitative description of individual solar cycles' features, such as onset, intensity, evolution, in terms of the number of M and X-class solar flares. The main elements of the model are the relative ecliptic motion of the planets Jupiter and Saturn, and its synergy with a quasi-periodic component of solar activity. Using as input the temporal distribution of flares during cycle 21, the general evolution of cycles 22-24 is reproduced in notable agreement with the observations, including the resurgence of activity in the last months of 2017, and further predictions are provided for cycle 25. This deterministic description could contribute to elucidating the responsible physical mechanisms and forecasting space weather.

## 1. Introduction

Energetic solar events and the quasi-periodic variability in solar activity, known as the solar cycle, are widely attributed to the Sun's magnetic dynamo mechanism (Parker, 1955; for a recent review, Brun and Browning, 2017); however their modelling is still far from complete (e.g. Spruit, 2010; Brun and Browning, 2017) and no regulating factors have been established. Existing methods for the prediction of the timing and amplitude of solar cycles mainly depend on extrapolations from sunspot numbers or geomagnetic precursors (e.g.

Hathaway et al., 1994, 1999), becoming available only very close to or after a cycle's start and often departing from the actual events (Usoskin and Mursula, 2003; Pesnell, 2008; Hathaway and Wilson, 2016; NOAA, 2009), although recently proxies like the solar background magnetic field enable new approaches (e.g. Zharkova et al., 2015). In the current article a deterministic model is presented for the quantitative description of the cycles' evolution, in terms of the number of M and X-class solar flares. Section 2 presents the used data and conventions; the derivation of the model and its results and predictions are presented in Section 3, with a brief discussion found in Section 4. A preliminary form of this work first appeared in February 2017 (Petrakou, 2017).

## 2. Data and conventions

The observable of choice in studies of the solar cycle has traditionally been sunspots, however the last four decades made possible the daily recordings of solar flares. While sunspots are indirect indicators of underlying dynamics, flares constitute actual physical events with definite timing and energy, as well as impact on space weather, and this study will focus on them. In the current article M-class and X-class flares (covering X-ray brightness of $10^{-5}$ W/m$^2$ and above) are used; only the counts of these flares are examined, treating them as statistical timed events, while less energetic flares which occur in large numbers almost daily are not included. However, the use of C-class flares and brightness is discussed towards the end of Section 3.

Solar flares data comprise the X-ray flux measurements of the NOAA SMS and GOES satellites and are provided by the USA National Oceanic and Atmospheric Administration (NOAA, 2017). Data on sunspots come from the archives of the Royal Observatory of Belgium (SILSO, 2017).

The presented model was developed using the data since the start of cycle 21 and up to the end of year 2016, in total 6,339 and 491 M and X-class flares, with the two categories corresponding to X-ray brightness of $10^{-5}$-$10^{-4}$ W/m$^2$ and all higher values, respectively. A corresponding definition is used for the cycle start and end (instead of the customary sunspot cycle). The start of each cycle is defined by the date of the first M-class flare erupting from a sunspot of reversed magnetic polarity (these flares are also the first ones after the minimum in flare activity, and they come after the minimum in sunspot activity, although in two of the cases they are not the first ones after the latter). The resulting start dates for cycles 21-24 are: 1977/01/31, 1986/10/19, 1997/04/01, 2010/01/19. The end of each cycle is defined by the start of the next one.

All quoted angles will refer to the relative heliocentric ecliptic longitude between Jupiter and Saturn (HelioWeb, 2017). With the exception of Fig.2, both conjunction and opposition are set equal to zero degrees, thus the range of values is [-90°, 90°]. Fig.1 illustrates three examples of the relative angle. In this convention, "91°" is actually -89°, since the closest alignment is the next opposition.

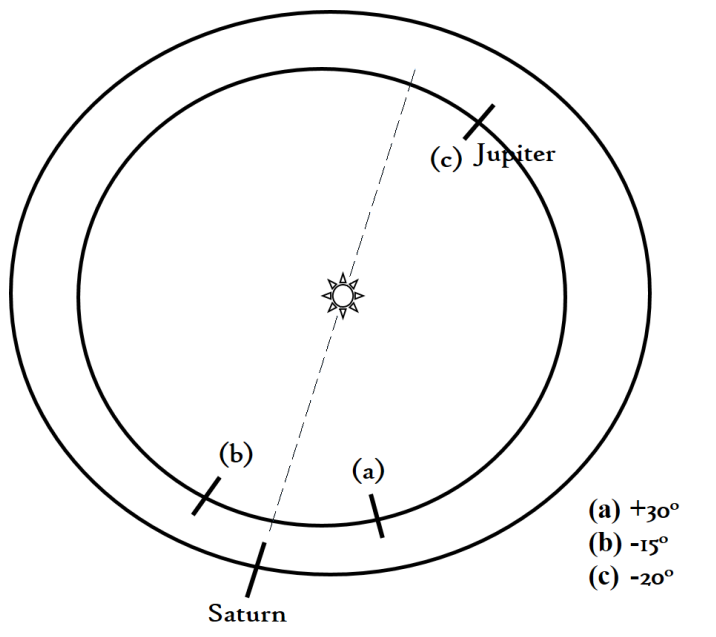

Figure 1: Examples of the planetary relative angle convention.

## 3. Observations and calculation

The model is initiated by the empirical observation that solar activity in terms of energetic flares tends to peak around the dates of alignment of Jupiter and Saturn, and be bound within the surrounding range roughly defined by the dates of their quadrature (Fig.2.a). However, as we progress from cycles 21 to 24 the activity is "dragged" further away from the alignment towards later dates (Fig.2.b). As this lagging is compatible with the staggering between the two planets' synodic period and the observed solar cycles duration of ∼11 years, it can be asked whether the evolution of solar activity is the coupled effect of two contributions: an internal mechanism generating the 11-year cycle, presumably of magnetic origin, and a triggering associated with the approach and retreat of Jupiter and Saturn.

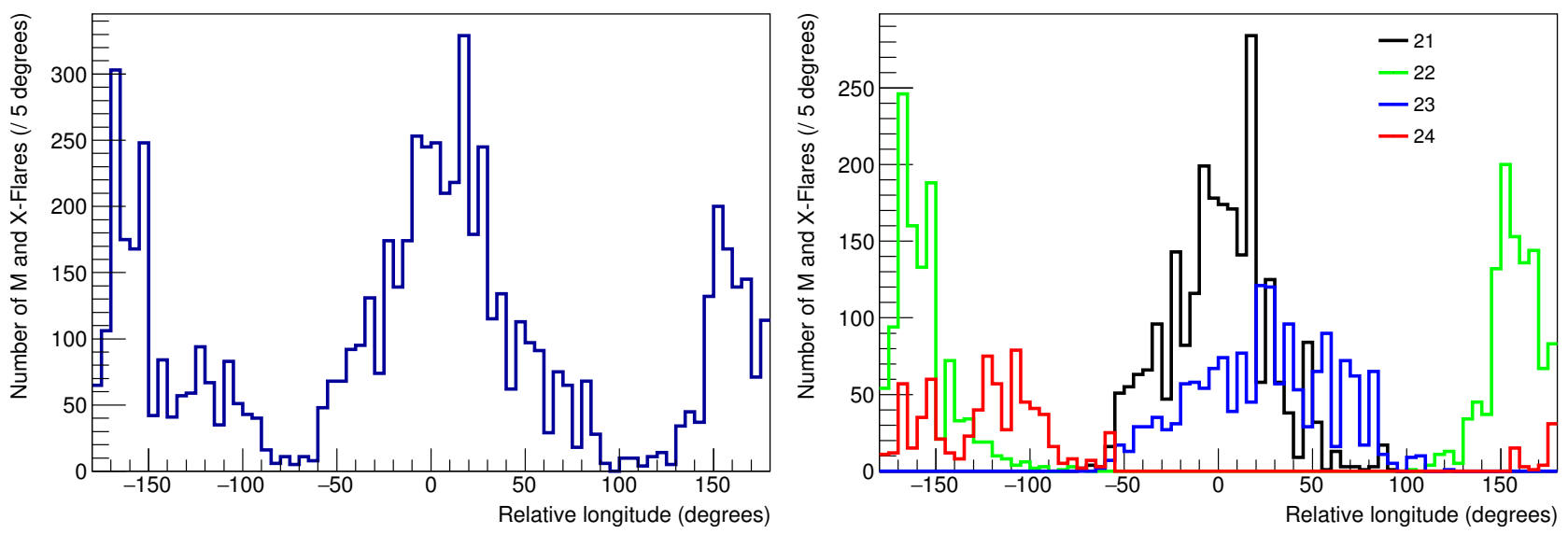

Figure 2: Solar flares – planetary angular relation. Number of M and X-class flares of cycles 21-24 up to the end of year 2017, as a function of the relative ecliptic longitude between Jupiter and Saturn, (a) collectively and (b) individually for each cycle (see text for definition of cycle start and end).

This proposition can be quantified by assuming that the effects of each contribution can be expressed by a Gaussian distribution with known mean and roughly known standard deviation: the distribution corresponding to the internal component would be centered on the temporal middles of cycles and span somewhat less than 11 years; and the distribution corresponding to the "Jupiter-Saturn component" would be centered on the dates of their alignments and lie mostly between -45$^o$ and +45$^o$ with respect to the alignments (the last requirement stemming empirically, Fig.2). Noting that in cycle 21 the dates of the temporal middle and of the alignment happened to lie close (237 days away), it will be assumed that during that cycle the full deployment of the two effects can be observed. This enables the extraction of the two distributions from the data of cycle 21, by finding two Gaussian functions which satisfy the described bounds for the mean and the standard deviation, and follow the envelope of the recorded activity within each component's respective time span (Fig.3). The two resulting functions' constants are close and they were refitted with equal values (fits performed with the ROOT package, Brun and Rademakers, 1997). The parameters of the two distributions are $\{\mu, \sigma, c\} = \{0, 670, 190\}, \{-237, 510, 190\}$.

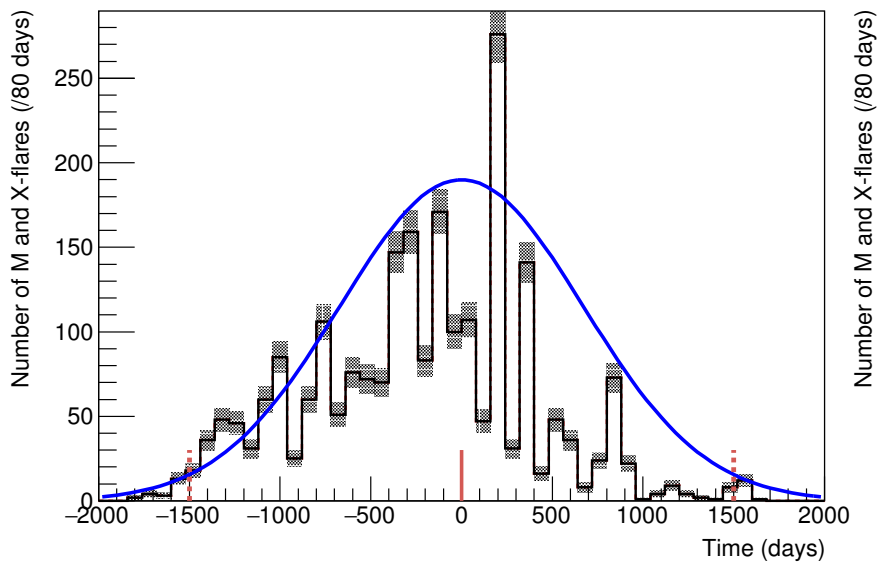

Figure 3: Extraction of the temporal "component distributions" from cycle 21. Number of flares in cycle 21 centered on its temporal middle, and the extracted Gaussian functions corresponding to the internal component (left) and the Jupiter-Saturn component (right). Guides are plotted at the respective centerings and fitting ranges; left: cycle's middle and $\pm$1,500 days; right: date of alignment and dates of $\pm 45^{\mathrm{o}}$. (Poisson errors are shown on the data to illustrate the ranges used by the fitting algorithm.)

By expanding over the time range of the latest four cycles and repeatedly placing the two distributions at the relevant dates, i.e. centering the Jupiter-Saturn distribution on the dates of alignments and the internal distribution on the temporal middles of cycles, Fig.4.a is obtained. On average, the distance between these two dates increases by 396 days between consecutive cycles (given the synodic half-period average of 3,634 and the sunspot cycle average of 4,030 days); this number was used for estimating the temporal middle for the ongoing cycle 24, with respect to 23. The model is completed by the assumption that the coupling of the two components is expressed by their common area, shown as a binned histogram in Fig.4.a. The assumptions used in this construction (the use of M and X-class flares, relevant cycle timing, the presence of two components, the modeling by Gaussian functions with the assumed span and timing, their extraction from cycle 21, their coupling) form the set of hypotheses to be tested against the data.

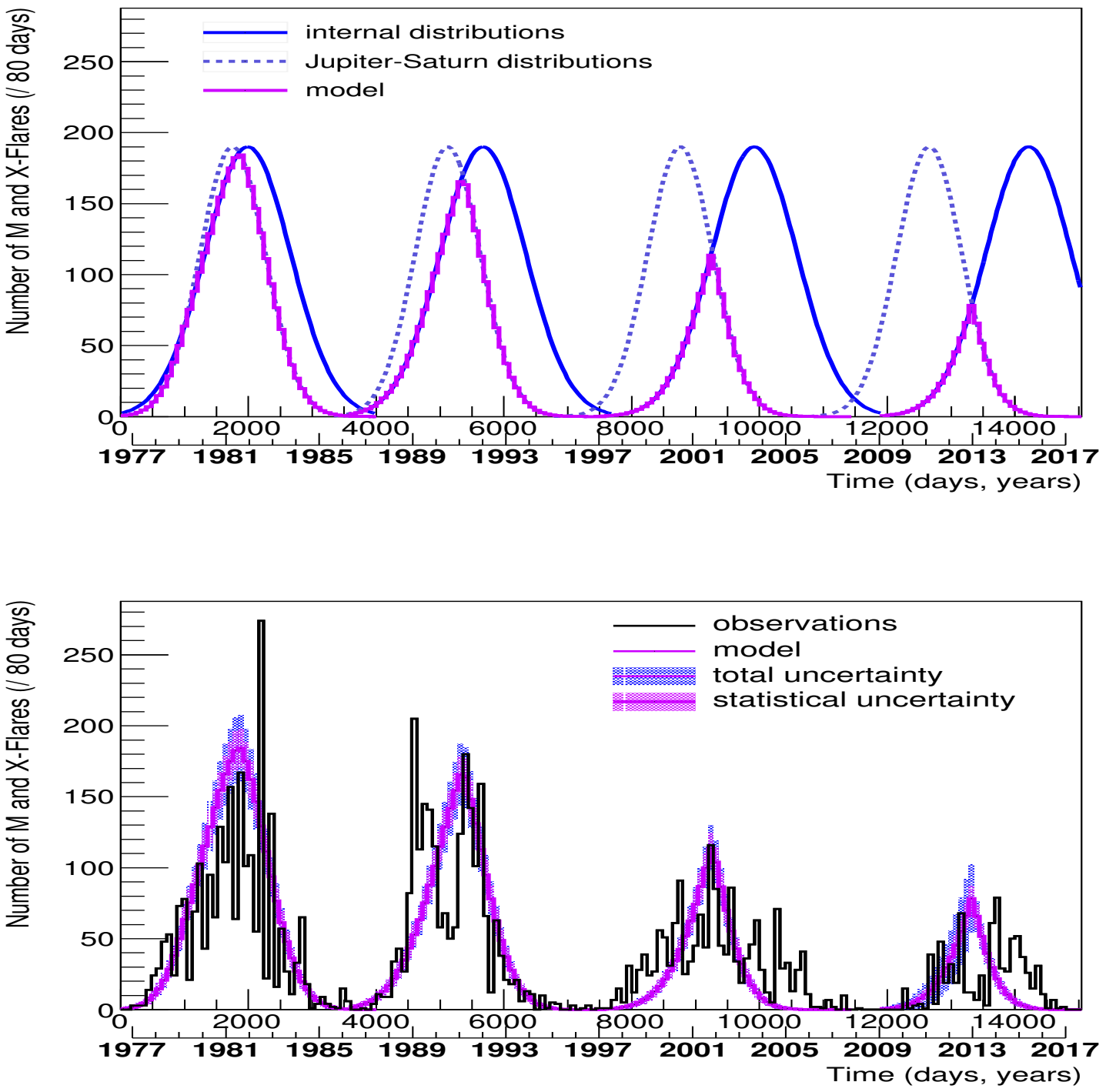

Figure 4: Model predictions and comparison to observations for cycles 21-24. (a) The two types of Gaussian components centered on cycle middles and alignment dates respectively, and the binned distributions resulting from their overlap. (b) Number of M and X-class flares of cycles 21-24 up to the end of year 2016, overlaid with the model distributions, including systematic and Poisson uncertainty. (The time range starts on 1976/06/30.)

The last distribution is proposed to describe the long-term solar activity in terms of energetic flares; in Fig.4.b it is overlayed with the observations up to the end of year 2016, including systematic uncertainties from the binning choices and from the timing of cycle 24 (Appendix A). Notable agreement can be seen in general features such as start and time span of activity, intensity, and evolution of each cycle. Short-term departures need to be understood in more depth, more prominent ones being the excess in the descending phase of the two latest cycles. However, certain short-term features which are generally consid-ered puzzling (e.g. Hathaway, 2015) are present in the model, such as the deep minimum and late onset of cycle 24, and the abrupt decrease in activity after the year 2015 (with the two planets retreating further than $+90^{o}$ in December 2015, marking the pause of activity before a new build-up begins with their approach).

For clarity, it can be pointed out that the width of the internal distribution is not meant to correspond to the duration of the sunspot cycle, or any other solar activity cycle, but is a measure of the span of the internal component's influence. Although the main coverage of these distributions is taken as fixed, their central dates vary to follow each cycle's individual duration. Likewise, the range of $-45^{o}$ to $+45^{o}$ reflects the main span of the planetary component's influence, while its centering on the dates of alignment follows the actual different time lengths between consecutive alignments. As discussed, the activity within each cycle is proposed to arise from the coupling of the two components (i.e. not by the internal component alone). The element which has to be accepted "as is" is the temporal middle of each cycle (however it is shown that it can be approximated to a satisfactory degree for one or two consecutive cycles, Appendix A).

The small number of available cycles might not lend itself to conclusions, yet at any rate the non-randomness of the presented model is supported statistically. The Pearson correlation coefficient between the distributions of flares from observations and from the model (Fig.4.b) has the values 0.73 and 0.88, in the whole range and within $0^{o}$ to $+45^{o}$ respectively, with $p$-values smaller than $10^{-7}$. Kolmogorov-Smirnov testing of the periodic displacement between the two used Gaussian distributions points strongly to a best result 1.3% away from the displacement's actual average value (Appendix B).

There is no obvious asymmetry in the distributions between the northern and southern hemispheres, with the exception of the two local maxima close to the predicted maximum counts for cycles 23 and 24, which are both "spikes" from the southern hemisphere. The inclusion of C-class flares "blurs" the shape of the distributions with the disproportionate number of low-energy events, but does not change qualitatively the picture. Furthermore, if the total brightness is examined instead of counts, the inclusion of all three classes results in distributions which follow satisfactorily the presented model, scaled appropriately (with the exception of isolated highly energetic events). Nevertheless, it is suggested that a detailed comparison between observables could possibly speak about different underlying effects. Another such plausible investigation would be on the influence of individual planets' motion, primarily Jupiter.

By repeating the use of the average increase in the displacement between the two Gaussian components, the model is extended over the next years (Fig.5). During both the ongoing and the next cycle, two Jupiter-Saturn Gaussians will overlap with the projected internal Gaussian, a case which did not occur in the three recorded whole cycles. However, although the core of the presented model is the common area of the two components, it is seen that activity also occurs outside that area in time ranges where both components remain substantial (Fig.4.b). Therefore, the double overlap can be reasonably expected to lead to a cycle 25 with spread-out activity characterised by two detached peaks, and in any case with intensity comparable to that of the current one.

Although this point about other time ranges is not quantified yet, it had resulted in the expectation of a surge in activity before cycle 24 finishes (Petrakou, 2017), as the two gas giants start approaching anew. This expectation is compatible with the activity in the second half of 2017 (Fig.5). The actual evolution of the rest of cycle 24 could offer clues about the similar case of cycle 25[1].

[1] An accurate comparison to the sunspot records is impeded by the absence of modeling for the correspondence between sunspots and flares, and the significant difference of their temporal distributions in cycles 21-24. However, it is perhaps noteworthy that if the presented model is extended to cover the range of the historical records, the only two times where this double overlap and the accompanying low distributions occur are in cycles 6 (Dalton minimum) and 14 (Gleissberg minimum).

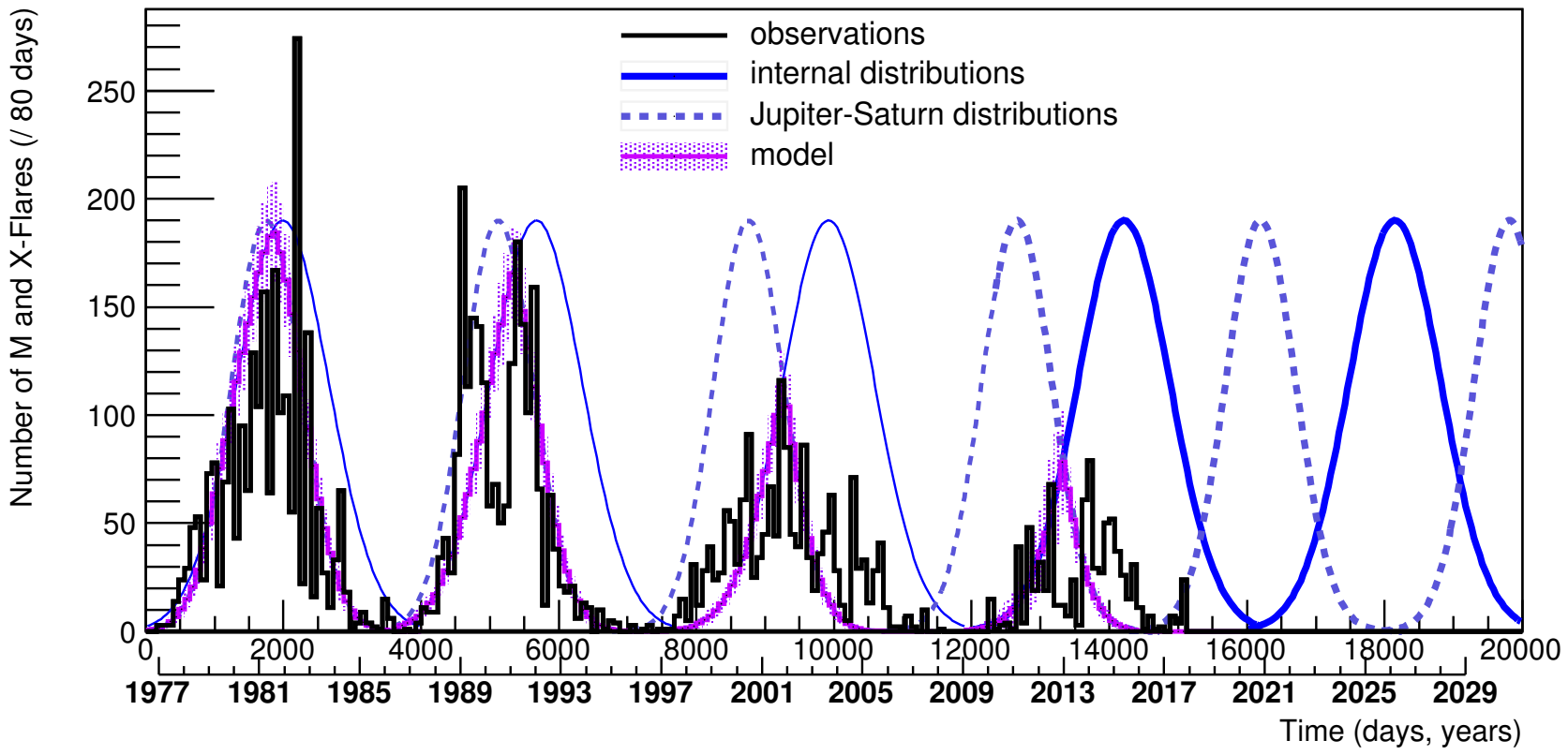

Figure 5: Observations and model predictions for cycles 21-25. Number of M and X-class flares of cycles 21-24 up to the end of 2017, overlaid with the model distributions. The two types of Gaussian components centered on the respective dates are also shown, highlighted since 2010. (The time range starts on 1976/06/30. After the last three populated bins, the next unpopulated bin starts on 2017/11/20.)

## 4. Discussion

Relations between planetary periods and solar activity have occasionally been pointed out over the past two centuries by a number of studies, however most of the relevant work is based on spectral or arithmetic analysis, with only a few attempts at quantification using timed solar events. Planetary tidal exertion on the Sun has been largely disfavoured, based on its magnitude (de Jager & Versteegh, 2005) and "unresponsive" stretches such as the Maunder minimum (Smythe and Eddy, 1977). However, amplifying mechanisms of the tidal effects have been proposed (Abreu et al., 2012; Scafetta, 2012a) and the sporadic absence of activity has been countered by higher resonances and coupling to internal activity (Abreu et al., 2012; Scafetta, 2012b), while other possible explanations for a planetary relation have been put forth. Although at this stage of the present work no inferences are drawn about possible underlying mechanisms, some notable recent studies can be listed indicatively. Abreu et al. (2012) examines a planetary torque exertion on the tachocline, and Sharp (2013) focuses on Uranus and Neptune and discusses a spin-orbit coupling mechanism based on the motion of the solar system's barycenter (Jose, 1965), with the latter also elaborated recently in Wilson et al. (2008) with findings of some similarity to those presented here. The spectral analysis in Scafetta (2012b) focuses on the combined effects of Jupiter and Saturn, employing three main harmonic components. (Particularly, the two Gaussian components of the model presented here are compatible with two of those harmonics, corresponding to the tidal spring period of the two planets and the mean solar cycle. The "additional" component, corresponding to Jupiter's sidereal period, leads to the prediction of a minimum of similar relative intensity but shifted timing with respect to the one predicted here.) Notably, Bertolucci et al. (2017), statistically, and Hung (2007), case-by-case, have reported relations between planetary positions and solar flares. Finally, a relevant older phenomenological study is Nelson (1952), which related planetary positions with terrestrial radio propagation conditions.

This article presented a model for the phenomenological description of long-term solar activity and the quantification of the main features of solar cycles in terms of energetic flares. Its principal element is a coupling between the empirical cyclic element of approximately 11 years and the relative ecliptic longitude of the planets Jupiter and Saturn, expressed by the common area of two distributions extracted from the observations of cycle 21. The sole other element required is the date of a cycle's temporal middle; obtaining it from the observations renders the model descriptive, while estimating it from the periodic increase in displacement between the two components leads to predictions. Thereby, using as input the observations of cycle 21, the distributions of energetic flares activity in the latest three cycles are reproduced to a notable degree. The model is extended to the next years, providing predictions for the rest of cycle 24 and cycle 25.

Although there is no suggestion made about the underlying physical mechanism, these results point to a correlation between the triggering of solar activity and the relative position of the gas giants, with the activity increasing and declining respectively with their approach and retreat. This work is expected to contribute to the understanding of the mechanisms involved in solar dynamics and to a long-term forecasting of space weather.

### Acknowledgments

The author is indebted to Juan-Carlos Algaba, Jonghee Yoo, Kalliopi Iordanidou and Jonathan Elkin for the useful discussions, and particularly to Konstantin Zioutas.

The provision of solar X-ray flux measurements by the USA National Oceanic and Atmospheric Administration and sunspots data by the Royal Observatory of Belgium is acknowledged.

Part of this work was supported by the Institute for Basic Science, Republic of Korea (Center for Axion and Precision Physics Research, grant IBS-R017-D1-2017-a00).

## Appendices

### Appendix A. Systematic uncertainties

Two sources of systematic uncertainty enter the calculations, from the binning choices and from the placement of the internal component's Gaussian dis-tribution.

Both the bin size and the position of the start date of data within the first bin modify the short-scale distribution of flares. The induced uncertainties are derived independently for each case, from the standard deviation of $\chi^2$ between the original model and the data of cycle 21, for 7 different binning choices. The resulting uncertainty is 10.4% on the original $\chi^2$ from the bin size and 7.0% from the start date. The larger value is applied to the model in all cycles.

For cycles 24 and 25, the internal Gaussian distribution was centered according to the average increase in its distance from the two planets' alignment date (Section 3). The resulting uncertainty is estimated from comparison of this method against the observed lengths of cycles 22 and 23; the projected cycle middles fall +81 and -211 days away from the actual ones, with their modulus average at 146 days. This uncertainty is applied to the centering of the Gaussian in the ongoing cycle 24.

The uncertainties are applied in quadrature with their effect shown in Fig.4.b.

### Appendix B. Kolmogorov-Smirnov testing

The Kolmogorov-Smirnov test is used to examine whether a satisfactory agreement between the data and the overlapping of the two used Gaussian distributions could arise randomly as a result of their staggering, or if it is indeed linked to the two planets' synodic period.

The average time length between consecutive Jupiter-Saturn alignments is 3,634 days. By letting this length vary arbitrarily, while keeping the timing of the internal components fixed, the model can be re-calculated for an arbitrary staggering between its two Gaussian components. Thus the length is iterated between 2,000 and 5,000, with a step of 30 days, and each resulting distribution is tested for compatibility to the observations. The compatibility is checked by applying the Kolmogorov-Smirnov test on the pair of predicted and observed distributions over the whole available time range, thus obtaining a value of the Kolmogorov-Smirnov distance for each step of the varied time length.

The minimum value of the distance indicates the time length which results in a distribution most compatible with the data. This occurs at a well-defined global minimum of 3,680 days, i.e. 1.3% away from the actual value (Fig.B.6). If one takes into account the natural oscillation of 2% in the synodic period, then the minimum falls well within the associated uncertainty. (The test was performed with binning in 80 days, as in the rest of the analysis; its repetition with bin width of one day, for the ideal condition of unbinned data, results in a similar distribution for the Kolmogorov-Smirnov distance.)

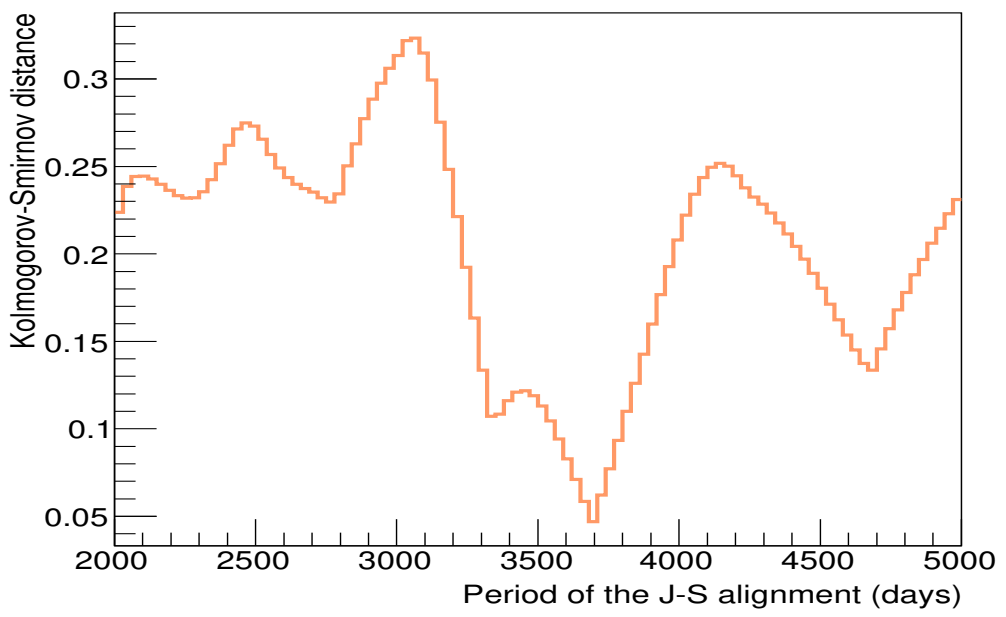

Figure B.6: Test of the compatibility of the observations to arbitrary staggering between the model's components. Kolmogorov-Smirnov distance for predictions with varying values of the average time length between the two planets' alignments, for binning in 80 days.